\documentclass[aps,prd,12point,twocolumn,nofootinbib,showpacs,superscriptaddress]{revtex4-2}
\def\theequation{\arabic{section}.\arabic{equation}}
\usepackage{psfrag}
\usepackage{subfigure}
\usepackage{color}
\usepackage{mathrsfs}
\usepackage{graphicx}
\usepackage{amssymb, bm}
\usepackage{amsmath, amsthm}
\usepackage{epstopdf}
\usepackage{hyperref}
\usepackage{enumerate}
\usepackage{longtable}

\usepackage{amsfonts}
\usepackage[T1]{fontenc}
\usepackage{bm}
\usepackage{amsmath}  
\usepackage{graphicx} 
\usepackage{color}
\usepackage{appendix}

\newcommand{\be}{\begin{equation}}
\newcommand{\ee}{\end{equation}}

\begin{document}
\def\theequation{\arabic{section}.\arabic{equation}} 

\title{Maximum force and cosmic censorship}

\author{Valerio Faraoni}
\email[]{vfaraoni@ubishops.ca}
\affiliation{Department of Physics and Astronomy, Bishop's University, 
2600 College Street, Sherbrooke, Qu\'ebec, 
Canada J1M~1Z7}



\begin{abstract}

Although the idea that there is a maximum force in nature seems untenable, 
we explore whether this concept can make sense in the restricted context 
of black holes. We discuss uniformly accelerated and cosmological black 
holes and we find that, although a maximum force acting on these black 
holes can in principle be introduced, this concept is rather tautological.

\end{abstract}

\pacs{}

\maketitle

\section{Introduction}
\label{sec:1}
\setcounter{equation}{0}

The idea that there exists a maximum value 
\be 
F_\text{max}=\frac{c^4}{4G} \label{Fmax}
\ee
to any physically attainable force (or tension)  was advanced by Schiller 
\cite{Schiller1997,Schiller2005,Schiller:2006cm}
and Gibbons \cite{Gibbons:2002iv} and further explored by several authors 
\cite{Schiller2005, Schiller:2006cm, Bruneton:2013ena, Barrow:2014cga, 
Good:2014uja, Barrow:2017atq, 
Ong:2018xna, Barrow:2020tky}. 
Extensions to include a 
cosmological constant \cite{Barrow:2014cga}, angular momentum 
\cite{Barrow:2017atq}, and a maximum force in Brans-Dicke gravity  
\cite{Barrow:2020tky} have also been discussed.

The key idea leading to the sharp factor $1/4$ in Eq.~(\ref{Fmax}) comes 
from the fact that there is a maximum deficit angle in the geometry of a 
cosmic string \cite{Gibbons:2002iv,Barrow:2014cga}.  A static cosmic 
string aligned with the $z$-axis is described by a locally flat spacetime 
with line element
\be
ds^2=-dt^2 +dr^2 +dz^2 +r^2 d\varphi^2
\ee
with $0\leq \varphi < 2\pi \left( 1-\frac{4G\mu}{c^2} \right) $ and with 
a conical singularity along the $z$-axis and a deficit angle
\be
\delta = \frac{8\pi G\mu}{c^2} \,,
\ee
where $\mu$ is the string tension. The range of the $\varphi$ coordinate 
in cylindrical coordinates $\left( t, r, \varphi, z \right)$ is $0 \leq 
\varphi \leq 2\pi (1-\delta)$.  To prevent $\delta$ from spanning the 
entire three-dimensional 
space it must be $\delta < 2\pi$, which generates the upper 
bound~(\ref{Fmax}) on the string tension $\mu$. It is not clear, however, 
how this bound on a string tension has risen to the role of a universal 
bound on any possible force acting on a particle.  

Sometimes the factor $1/4$ is replaced by $1/2$, or by factors of order 
unity, in weaker formulations of 
the upper bound \cite{Ong:2018xna}, while other works involving black 
holes reproduce the $1/4$ factor \cite{Good:2014uja}.

Another argument involves the Planck scale. Using only dimensional 
analysis and the fundamental constants $G,c$, and $\hbar$, one can 
construct the ``Planck force'' $F_\text{Pl}=c^4/G$ and the ``Planck 
power'' 
$P_\text{Pl}= c^5/G$. 
Contrary 
to the Planck length, energy, temperature, and energy 
density, these two new Planck scale quantities do not contain $\hbar$ and 
are purely classical. The construction of Planck scale quantities, 
however, may lead to very vague concepts: for example, ordinary objects 
like a football have masses exceeding the Planck mass and, therefore, a 
force resulting   from the product of a mass times an acceleration could 
easily exceed the Planck force even if there was an upper limit to the 
acceleration of  a particle (such a limit was claimed by Caianiello on the 
basis of a generalized uncertainty principle \cite{Caianiello81}).

The idea of a maximum force has been extended to hypothesizing a maximum 
power for any system in general relativity (GR), the Dyson luminosity 
\cite{Dyson}
\be
P_\text{max} = cF_\text{max} = \frac{c^5}{4G} \,,
\ee
which would be the maximum possible luminosity of an isolated system 
\cite{Schiller1997, Schiller:2006cm} (for example, its luminosity in 
gravitational waves \cite{Sperhake:2011xk,Cardoso:2013krh}).

Schiller's maximum force proposal  in \cite{Schiller1997} (see also  
\cite{Schiller2005, Schiller:2006cm}) even included the 
idea that the existence of a maximum force implies general relativity 
as a consequence, in the same way that a maximum possible velocity ($c$) 
leads 
to special relativity \cite{Schiller2005}. The derivation parallels 
Jacobson's derivation of the Einstein equations as an equation of state 
that begun the area of research now known as the thermodynamics 
of spacetime \cite{Jacobson:1995ab}. 
This derivation is associated with the fact that the Einstein equations
\be
G_{ab}=\frac{8\pi G}{c^4} \, T_{ab}
\ee
contain the constant $G/c^4$ and can be written as  
$T_{ab}=\frac{F_\text{max}}{2\pi} \, G_{ab} $ \cite{Jowsey:2021ixg}.

Recently, the existence of a maximum force has been criticized by Jowsey 
and Visser \cite{Jowsey:2021ixg} who provide counterexamples violating the 
Schiller-Gibbons upper bound using fluid spheres, or the TOV equation, in 
GR. Counterexamples to the maximum luminosity conjecture are given 
by Cardoso, Ikeda, Moore and Yoo \cite{Cardoso:2018nkg}. 

Indeed, Barrow 
and Gibbons themselves mention possible counterexamples related with 
sudden future singularities in cosmology, which cause unbounded pressure 
forces. They mention the fact that the maximum force bound is restricted 
to situations in which a suitable energy condition is imposed to prevent 
sudden future singularities \cite{Barrow:2014cga}. However, this example 
is rather vague and it is not clear what the cosmological ``force'' 
applies to. Sec. ~1.5 of \cite{Barrow:2014cga} begins with the usual 
Newtonian analogy for an expanding universe, consisting of a Newtonian 
ball 
of material with radius proportional to the scale factor $a(t)$ of a 
(spatially flat) expanding universe. Since the ``force'' is proportional 
to $\ddot{a}$, assuming a power-law expansion $a(t) \simeq t^n$, the 
``force'' $F$ grows like
\be
F = F_\text{Pl} \left( \frac{t}{t_\text{Pl} } \right)^{n-2} \,.
\ee
If $n>2$ the Planck force is exceeded. In our opinion, however, this 
Newtonian argument does not suffice to draw firm conclusions. In fact, the 
only way to make the analogy with a Newtonian ball work in GR is if the 
analogous relativistic universe is filled with dust 
\cite{Faraoni:2020uuf}. Then it is $n=2/3$ and the ``force'' decreases 
after the Planck time. This entire argument, however, is too vague and 
needs to be put on a firmer basis (in particular specifying what is the 
force acting on what).

Varying constants and entropic force theories, as well as black hole 
thermodynamics in generalized entropy models also escape the force 
bound \cite{Dabrowski:2015eea}, while attempts have been made to 
relate it to the holographic principle \cite{Bolotin:2015uwa}.

On the one hand, in view of the counterxamples provided, it is hard to 
argue with Jowsey and Visser that the maximum force acting on particles, 
in GR or in other theories of gravity, is doomed. On the other hand, 
Gibbon's argument in the restricted context of cosmic strings is sound and 
perhaps versions of the maximum force conjecture restricted to well 
defined contexts may be true. Here we explore this direction:  although 
there is no universal maximum force on particles and other objects, there 
may exist an upper bound to the force (or to a similar quantity with the 
dimensions of a force) acting on more fundamental objects: {\em black 
holes}. We report two instances that corroborate this idea. They include 
i)~uniformly accelerated black holes and ii)~black holes embedded in 
Friedmann-Lema\^itre-Robertson-Walker (FLRW) universes and described by 
the McVittie metric \cite{McVittie}. For these two categories of black 
holes, we present first a 
purely classical argument and then a second argument based on horizon 
thermodynamics (therefore, ultimately based on quantum field theory in 
curved space).

We follow the notation of Ref.~\cite{Waldbook}.

\section{Uniformly accelerated black hole}
\label{sec:2}
\setcounter{equation}{0}

The static C-metric discovered by Levi-Civita in 1917 
\cite{LeviCivita1917}, a member of the Weyl class of 
cylindrically symmetric solutions of the Einstein 
equations \cite{Weyl1919}, 
describes a pair of uniformly accelerated black holes and has been 
generalized to include electrically and magnetically charged and/or 
spinning black holes, possibly with a positive or negative cosmological 
constant. The C-metric has been the subject of a long literature 
\cite{Cmetric}. It can be studied in several coordinate systems, each one 
of which is more suitable for certain purposes than others 
\cite{Griffiths:2006tk}.

The C-metric can be written in coordinates $\left( t, p, q, \varphi 
\right)$ and in units in which $c=G=1$ as the 2-parameter 
family of solutions  \cite{FarhooshZimmerman80}
\be
ds^2 = \frac{1}{a^2 \left(p+q\right)^2} \left[ -F(q)dt^2 
+\frac{dq^2}{F(q)} 
+\frac{dp^2}{G(p)} +G(p) d\varphi^2 \right] \,, \label{Cmetric1}
\ee
where
\begin{eqnarray}
F(q) &=& -2amq^3 +q^2-1 \,,\\
&&\nonumber\\
G(p) &=& -2amp^3 -p^2 +1 \,.
\end{eqnarray}
The parameters $m$ and $a$ are related with the black hole mass 
and the uniform acceleration. The coordinates have the range $-\infty < t< 
+\infty$, $0\leq \varphi \leq 2\pi$, while $p$ and $q$ have different 
ranges chosen to satisfy $G(p)>0$. The cubic polynomial $G(x)=-F(-x)$ has 
three real roots if $am<1/\sqrt{27}$ \cite{FarhooshZimmerman80}. If 
$q_\text{R,S}$ and $p_{0,\pi}$ denote the roots of $F(q)=0$ and $G(p)=0$,  
it is $ q_\text{R} \leq q \leq q_\text{S} $ and  $ p_0\leq p\leq p_{\pi}$. 

In the limit $m\rightarrow 0$, the line element~(\ref{Cmetric1}) reduces 
to the Minkowski   one in Rindler accelerated coordinates, which allows 
the identification of the parameter $a$ with the uniform acceleration 
\cite{FarhooshZimmerman80}.  
The limit $a\rightarrow 0$ reproduces the Schwarschild metric, but a  
transformation to different coordinates that reduce to the usual 
Schwarzschild ones when $a=0$ is necessary to see that 
\cite{Griffiths:2006tk}.   The metric is of algebraic type D, has a 
spacetime singularity inside the black hole horizon, and is static and 
cylindrically symmetric since it admits a timelike Killing vector $\xi_t$  
and a 
rotational Killing vector $\xi_{\varphi}$, plus a boost vector 
\cite{FarhooshZimmerman80}. The 
timelike and boost Killing vectors generate null Killing horizons, located  
by the vanishing of their norms. The first is a black hole horizon 
distorted by the acceleration, while 
the second is an acceleration (Rindler) horizon distorted by the presence  
of the black hole. These Killing horizons are best studied in 
the accelerated coordinates associated with observers comoving with the 
uniformly accelerated black hole \cite{FarhooshZimmerman80}. If  
$ma<1/(3\sqrt{3})$, the black hole and the Rindler horizon are separated, 
the black hole horizon is elongated along the direction of its motion,  
and the Rindler horizon is an infinite open surface around the black hole 
horizon 
(most of the literature on accelerated black holes restricts to this 
situation \cite{Cmetric}). A second black hole, joined to the first by a  
cosmic string, is located in the 
region behind the Rindler horizon \cite{Cmetric}. 
If $ma=1/(3\sqrt{3})$, the two horizons touch each other while, if 
$ma>1/(3\sqrt{3})$ the Rindler horizon penetrates the black hole horizon 
and the black hole effectively disappears from this region of spacetime. 

The geometry has a conical singularity along the axis corresponding to the 
direction of motion and to a cosmic string, aligned with this axis, that 
pulls the black hole. There is a deficit angle $\delta$ that manifests 
itself  when the length of a circumference in a plane orthogonal to this 
axis is computed for $0\leq \varphi < 2\pi$.  The string tension is $\mu = 
\frac{c^4}{G} \, \frac{\delta}{8\pi} $; imposing Gibbons' argument  
 \cite{Gibbons:2002iv} that this deficit angle be less than $2\pi$ yields 
$\mu \leq \frac{c^4}{4G}$. This bound is not the same as the bound on 
$am$; however, $m$ cannot be interpreted literally as the black hole mass 
since the spacetime is not asymptotically flat and the mass parameter is 
redefined in a complicated way \cite{Hong:2003gx} when the coordinates are 
transformed. In any case, the quantity $ma$ no doubt has the dimensions of 
a force acting on the black hole horizon, although the physical 
interpretation of the C-metric compels one to regard the string tension 
as the force acting on the black hole.

From the point of view of accelerated observers comoving with the black 
hole, the black hole is destroyed ({\em i.e.}, it disappears from these 
observers' world) if the parameter $a$ becomes too large.  The condition 
$ma\leq \frac{c^4}{3\sqrt{3}\, G}$ can also be obtained naively by 
remembering that the distance of a uniformly accelerated observer to the 
Rindler horizon is $d\leq \frac{c^2}{2a}$, which implies that the size of 
a uniformly accelerated object must be not larger\footnote{Incidentally, 
by imposing that the Compton wavelength $\lambda=h/(mc)$ of a particle of 
mass $m$ lies outside its Schwarzschild radius, one obtains the bound 
$a\leq a_c(m)/(8\pi)$ on the acceleration $a$, where $a_c (m) \equiv 
2mc^3/h$ is Caianiello's maximal acceleration \cite{Caianiello81}. (Loose 
arguments like this are very unlikely to produce exactly the same 
numerical factors.)  This argument, however, brings in quantum physics.} 
than $c^2/(2a)$. When the ``object'' is a Schwarzschild black hole of 
radius $R_S=2Gm/c^2$, the bound $R_S\leq c^2/(2a)$ reproduces $ma \leq 
\frac{c^4}{4G}$. This is, however, rather hand-waving while the derivation 
using the Farhoosh-Zimmerman solution is exact.

A thermodynamical argument can be given. The black hole and acceleration 
horizons are distorted,\footnote{See Refs.~\cite{Appels:2016uha, 
Astorino:2016ybm, Appels:2017xoe, Gregory:2017ogk, Anabalon:2018ydc, 
Anabalon:2018qfv, Gregory:2019dtq} for the exact thermodynamics of 
accelerating black holes in anti-de Sitter space.} however one can still 
approximate their temperatures with the quantities pertaining to an 
isolated black hole and a Rindler horizon in Minkowski space. The result 
is not rigorous, but is very suggestive.

The Unruh temperature of the thermal bath perceived  by a uniformly 
accelerated observer is
\be
T_\text{U}= \frac{\hbar a}{2\pi K_Bc} \,,
\ee
while the Hawking temperature of a Schwarzschild black hole of mass $m$ is
\be
T_\text{H}= \frac{\hbar c^3}{8\pi G K_B m} \,.
\ee
The requirement $T_\text{U} \leq T_\text{H}$ translates to
\be
ma \leq \frac{c^4}{4G} \,,
\ee
which is exactly the maximum force proposed by Gibbons and Schiller for a 
particle of mass $m$. However, in their proposals, the particle is not  a 
black hole and the acceleration $a$ is not uniform.

A physical interpretation of this argument can be given as follows. The 
typical quanta of Hawking radiation from the black hole have wavelength 
$\lambda \sim R_s$. When $T_\text{U}>T_\text{H}$, this wavelength is 
larger than the 
acceleration horizon and one can no longer talk about thermal emission. 
This is consistent, of course, with the fact that when the black hole 
horizon becomes larger than the acceleration horizon, it does not make 
sense to talk about a black hole. This echoes a situation discussed by 
Barrow \cite{Barrow:2002mj} in the independent context of varying speed of 
light cosmologies, in which the Compton wavelength of a particle crosses 
outside the particle horizon of the universe.

The thermodynamical argument is conceptually different from, and 
independent of, Gibbons' argument based on the deficit angle caused by the 
cosmic string pulling the black hole. It is interesting that it provides 
the same bound, although the coefficient of $c^4/G$ cannot be taken 
literally because the Unruh and Hawking temperatures employed are 
approximations. In the next section we examine a physical situation in 
which black hole horizons appear in the complete absence of cosmic 
strings, but similar considerations ensue.

\section{Black holes embedded in FLRW universes}
\label{sec:3}
\setcounter{equation}{0}

The McVittie spacetime \cite{McVittie} generalizes the Schwarzschild-de 
Sitter (or Kottler) solution of the Einstein equations and is interpreted 
as describing a central object embedded in a generic FLRW space 
\cite{Sussman85, Krasinskibook, NolanPRD, NolanCQG, Nolan2, 
KaloperKlebanMartin10, Roshina1, Roshina2, LakeAbdelqader11, 
SilvaFontaniniGuariento12, GuarientoAfshordi1, 
GuarientoAfshordi2,Faraoni:2015ula}. The geometry in the region between 
black hole and cosmological apparent horizons is time-dependent.

The McVittie line element in isotropic coordinates  is
\begin{eqnarray} 
ds^2 &=& -\frac{  \left[ 1-\frac{m_0}{2\bar{r}a(t)} 
\right]^2}{
\left[ 1+\frac{m_0}{2\bar{r}a(t)} \right]^2} \, dt^2+
a^2(t) \nonumber\\
&&\nonumber\\
&\, & \times
\left[ 1+\frac{m_0}{2\bar{r}a(t)} \right]^4 \left( 
d\bar{r}^2 +\bar{r}^2 d\Omega_{(2)}^2 \right) \,.
\end{eqnarray}
Apart from the special case of a  de 
Sitter ``background'' (in which McVittie reduces to Schwarzschild-de 
Sitter), there is  a spacetime singularity at $  \bar{r}=m/2 
$ (which  reduces to the 
Schwarzschild horizon if $ a \equiv 1$)  
\cite{Ferrarisetal, NolanPRD, NolanCQG, Nolan2, Sussman85}, which is 
spacelike \cite{NolanPRD,NolanCQG, Nolan2}. The pressure of the fluid 
source  
\begin{equation} \label{pressure}
P=-\, \frac{1}{8\pi} \left[ 3H^2+\frac{2\dot{H}\left( 
1+\frac{m}{2\bar{r}} \right)  }{1-\frac{m}{2\bar{r}} }\right]
\end{equation}
diverges at $  \bar{r}=m/2 $ together with  the 
Ricci scalar $ {\cal R}=8\pi \left( \rho -3P \right) 
$ \cite{Sussman85,Ferrarisetal, NolanPRD, NolanCQG, 
Nolan2, McClureDyer06CQG, McClureDyerGRG}.

The areal radius is
\be
R \equiv a(t) \bar{r} \left( 1+\frac{m}{2\bar{r}} \right)^2 
\,;
\ee
restricting to a spatially flat FLRW universe  
for simplicity, the apparent 
horizons of the  McVittie metric are the roots of the equation $\nabla^c R 
\nabla_c R=0$, or 
\be\label{8}
H^2(t) \, R^3 -R+2m_0 =0 \,,
\ee
which is familiar from the Schwarzschild-de Sitter case if $H=$~const. 
Since here $H=H(t)$, the apparent horizon radii are time-dependent.  
There are two real and positive roots  $ R_1(t)$ and $R_2(t)$ if
$  m_0 H(t)<1/(3\sqrt{3}) $. To fix the ideas, consider  a dust-dominated 
FLRW universe with $ H(t)=2/(3t)$. Then there is a critical time  
$ t_* =2\sqrt{3} \, m_0 $  at which $ m_0 H(t)=1/(3\sqrt{3}) $. We can 
distinguish three situations occurring during the history of the McVittie 
universe: 

\begin{itemize}

\item At early times $ t<t_*$ we have $ m_0 >\frac{1}{3\sqrt{3} \,H(t)} $, 
$ R_1(t) $ and $ R_2(t) $ are complex and there are no apparent horizons.

\item At  $t= t_*$ we have 
$ m_0 =\frac{1}{3\sqrt{3}\,H(t)} $ and two coincident apparent horizons  
$ R_1=R_2=\frac{1}{\sqrt{3}\,H(t)} $ appear simultaneously.
 
\item As $ t>t_*$ it is $ m_0 < \frac{1}{3\sqrt{3}\,H(t)} $ and there are 
two distinct apparent horizons $ R_{1,2}(t) $: a black hole and a 
cosmological horizon.

\end{itemize}

The physical interpretation is that at late times the black hole fits 
inside the cosmological horizon and can properly be called a black hole, 
while this is impossible as $t\leq t_*$, when the ``force'' caused by the 
attraction of cosmic matter  enlarges the black hole horizon.


Instead of considering a ``force'' acting on the black hole apparent 
horizon and stretching it, we can consider a quantity with the dimensions 
of a force, {\em i.e.}, the product $Gm/c^2 \cdot H/c$ that measures the 
ratio of the sizes of the black hole apparent horizon $(\sim Gm/c^2 $) and 
the cosmological apparent horizon ($\sim c/H$). The black hole apparent 
horizon is smaller than the cosmological apparent horizon, or touches it, 
when 
\be
mH \leq \frac{c^4}{3\sqrt{3}\, G} 
\ee
If $mH> c^4/\left( 3\sqrt{3}\, G\right) $, the black hole horizon is 
outside the would-be 
cosmological horizon and the comoving observers of the underlying FLRW 
cosmology do not see a black hole at all.

One can consider again the corresponding thermodynamics. Approximating the 
temperature of the black hole with that of a Schwarzschild black hole of 
the same mass $m$, $T_\text{H} =\frac{ \hbar c^3}{8\pi G K_B m}$, and the 
temperature of the FLRW apparent horizon with the Gibbons-Hawking 
temperature (originally obtained for de Sitter space with $H=$~const. 
\cite{Gibbons:1977mu}), $T_\text{GH}=\frac{ \hbar H}{2\pi K_B c}$, the 
requirement that $T_\text{GH} \leq T_\text{H}$ is equivalent to
\be
mH \leq \frac{c^4}{4G} 
\ee
which is, again, the upper bound proposed by Gibbons and Schiller. The 
coincidence of the numerical coefficients of $c^4/G$ is, again, not 
significant because the black hole and cosmological horizons are modified 
with respect to the GR solutions containing only one or the other, and so 
will their temperatures. The following physical interpretation can be 
given: when the wavelength of the thermal Hawking radiation emitted by the 
black hole becomes larger than the cosmological apparent horizon, 
corresponding to $T_\text{GH} > T_\text{H}$, the concept of Hawking 
radiation 
loses 
meaning, consistent with the fact that the black hole itself is no longer 
seen by observers comoving with the FLRW cosmic fluid.  Particles beyond 
the Hubble horizon cannot fall into the black hole singularity.

\section{Conclusions and outlooks}
\label{sec:4}
\setcounter{equation}{0}

In both cases the Schwarzschild null event horizon is altered by the 
black hole environment: in the first case, it is distorted by the pull of 
the cosmic string causing the black hole to accelerate while, in the 
second  case, the horizon remains spherical but it becomes a 
time-dependent apparent horizon due to the gravitational pull of the 
cosmic. It is 
enlarged with respect to the Schwarzschild radius of a black hole with the 
same mass (the central singularity is also stretched to a finite radius 
singularity \cite{Faraoni:2015ula}).

Technically, in our two examples, the maximum force principle is respected 
by black holes in the sense that when it is violated the spacetime regions 
under consideration are no longer black holes according to the observers 
located in them. However, both examples are legitimate solutions of the 
Einstein equations also in the spacetime regions beyond the acceleration 
horizon (in the first case) or the cosmological apparent horizon (in the 
second case) and, as such, the maximum force principle is violated. 
Alternatively, there is no {\em a priori} reason why the Unruh temperature 
of an accelerated black hole cannot be larger than its Hawking 
temperature. According to Wien's law of displacement, $\lambda T =b$ for a 
blackbody, where $b$ is constant and $\lambda$ is the wavelength 
corresponding to the maximum of the blackbody spectral energy density 
(taken here as a typical wavelength).  Then, the situation $T_\text{U} 
>T_\text{H}$ 
corresponds to the typical wavelength $\lambda_\text{U} =b/T_\text{U}$ of 
Unruh quanta 
being larger than the typical wavelength $\lambda_\text{H}$ of Hawking 
quanta. 
However it seems that, as long as observers can undisputably establish the 
existence of a black hole horizon, a restricted maximum force principle 
limited to these horizons could be valid. This conclusion is not 
surprising: although no universal force limits exists, the very fact that 
observers are required to see a black hole imposes an upper bound on the 
force (rather loosely defined) acting on this black hole horizon. In both 
cases considered, the solution of the Einstein 
equations is still mathematically admissible, but it contains a naked 
singularity. This statement is rather tautological, as we have decided to 
consider a situation in which the black hole horizon exists, but is  
located beyond the Rindler horizon or the cosmological apparent horizon. 
Imposing that the black hole horizon is not removed and that the 
singularity inside it remains invisible to what are deemed to be 
physically relevant observers located outside the black hole horizon 
amounts to limiting the force acting on the black hole horizon. The lesson 
seems to be that the existence a maximal force acting {\em on black hole 
horizons} is (a bit tautologically)  tied to cosmic censorship. The 
reference to forces acting on event, apparent, or Killing horizons, 
however, establishes a very special and restricted context within which to 
talk about maximal forces and in no way implies the existence of universal 
upper bound on the forces acting on (classical or quantum) particles or 
bodies, in agreement with the conclusions of Ref.~\cite{Jowsey:2021ixg}.

\begin{acknowledgments} 

This article is dedicated to the memory of John Barrow, with whom we had a 
discussion on this subject. This work is supported, in part, by the 
Natural Sciences \& Engineering Research Council of Canada 
(Grant~2016-03803) and by Bishop's University.

\end{acknowledgments}


\end{document}